\documentclass[aps,prl,reprint,superscriptaddress]{revtex4-2}
\pdfoutput=1
\usepackage{color,hyperref}
\usepackage{graphicx}
\usepackage{amsmath}
\usepackage{amssymb}
\usepackage{mathrsfs}
\usepackage{xcolor}
\usepackage{mathtools}
\usepackage{gensymb}

\begin{document}
	
	\title{Polarization Enhanced Deep Optical Dipole Trapping of $\Lambda$-cooled Polar Molecules}

	\author{Thomas K. Langin}
	\affiliation{Department of Physics, Yale University, New Haven, Connecticut, CT, 06520, USA}
	\affiliation{Yale Quantum Institute, Yale University, New Haven, Connecticut 06520, USA}
	\author{Varun Jorapur}
	\affiliation{Department of Physics, Yale University, New Haven, Connecticut, CT, 06520, USA}
	\affiliation{Yale Quantum Institute, Yale University, New Haven, Connecticut 06520, USA}
	\author{Yuqi Zhu}
	\affiliation{Department of Physics, Yale University, New Haven, Connecticut, CT, 06520, USA}
	\affiliation{Yale Quantum Institute, Yale University, New Haven, Connecticut 06520, USA}
	\author{Qian Wang}
	\affiliation{Department of Physics, Yale University, New Haven, Connecticut, CT, 06520, USA}
	\affiliation{Yale Quantum Institute, Yale University, New Haven, Connecticut 06520, USA}
	\author{David DeMille}
	\affiliation{Department of Physics, University of Chicago, Chicago, Illinois 60637, USA}

	\date{\today}
	
	\begin{abstract}
		We demonstrate loading of SrF molecules into an optical dipole trap (ODT) via in-trap $\Lambda$-cooling.  We find that this cooling can be optimized by a proper choice of relative ODT and $\Lambda$ beam polarizations.  In this optimized configuration, we observe molecules with temperatures as low as 14(1)\,$\mu$K in traps with depths up to 570\,$\mu$K.  With optimized parameters, we transfer $\!\sim\!5$\% of molecules from our radio-frequency magneto-optical trap into the ODT, at a density of $\sim\!2\times 10^{9}$\,cm$^{-3}$, a phase space density of $\sim\!2\times 10^{-7}$, and with a trap lifetime of $\!\sim\!1$\,s.
	\end{abstract}
	
	\maketitle

	Ultracold molecular gases can be produced either by assembly from ultracold atoms~\cite{noy2008,kgc2014,pwz2015,gzw2016,rsj2017}, recently leading to the first demonstration of a quantum degenerate molecular gas~\cite{dvy2019}, or by direct cooling and trapping of molecules.  Recent progress on the latter path includes demonstrations of magneto-optical trapping (MOT)~\cite{bmd2014,nmd2016,aad2017,cdy2018}, sub-Doppler cooling~\cite{twt2017,dwy2020}, loading into conservative magnetic quadrupole~\cite{msd2018,wct2018} and optical dipole traps (ODTs)~\cite{cad2018,aad2018}, and observation of molecule-molecule~\cite{cad2020,abd2021,acd2019} and molecule-atom~\cite{jct2021} collisions.  These improvements on both paths to cold molecules bring us closer to realizing their potential as platforms for quantum simulation~\cite{cmy2009,mbz2006,psz2008,bdz2007}, quantum information~\cite{dem2002,ykc2006}, and precision measurement~\cite{lah2018,acme2018,cgc2017}.  
	
	These applications all require molecular gases with high density and low entropy, i.e., in or near the regime of quantum degeneracy.  Achieving this in directly cooled molecular species will likely require evaporative cooling, either of atoms that are co-loaded with molecules in a sympathetic cooling approach~\cite{mbw1997,bge2001,spj2020} or by using just the molecular species~\cite{vmy2020}.  This requires achieving high enough density that rethermalizing collisions occur rapidly.  Increasing the initial phase-space density is also desirable, as starting closer to unity will minimize the loss of molecules during collisional cooling.  

	\begin{figure}
		\includegraphics{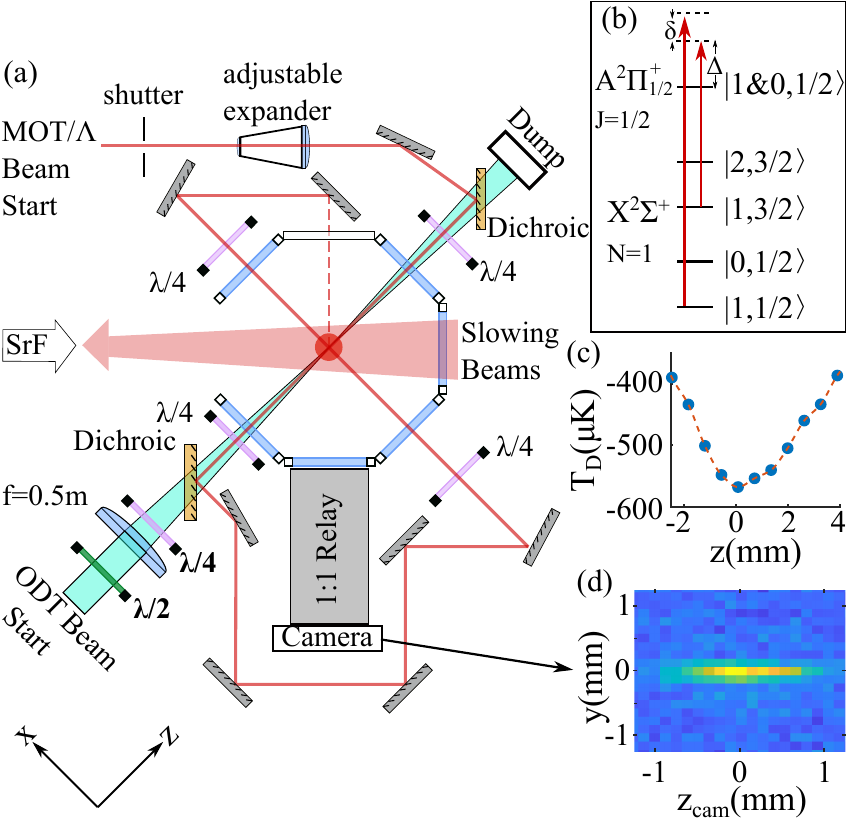}%
		\caption{(a) Experimental schematic.  The MOT/$\Lambda$-cooling beams pass through a shutter and an adjustable beam expanding telescope, then make three passes through the chamber before they are retroreflected along the same total path (dashed line indicates that the beam is directed down below the chamber, then reflected upward). The optical dipole trap (ODT) laser beam is combined with the $\Lambda$ beam using a dichroic mirror.  Waveplates before the dichroic (bottom left, bold labels) are used to control the polarization of the trap light.  (b) Level diagram for the $\Lambda$-cooling transition, with hyperfine sublevels $|F,J\rangle$ indicated.  (c) Trap depth vs.~position along axial dimension ($z$) of ODT.  (d) \textit{in-situ} image of optically trapped molecules.  Our imaging system does not resolve the width of the cloud along the radial dimension of the ODT (here, vertical).  However, the width of the molecular cloud along the horizontal camera axis ($\hat{z}_{cam}\propto(\hat{x}+\hat{z})$) is resolved.  We use the extent of the cloud along this axis, along with the trap profile in (c), to measure the temperature.  \label{fig:ODTSchematic}}
	\end{figure}
	
	In this Letter, we demonstrate the ability to reduce the temperature, $T$, (and thus maximize density, $n$, and phase space density $\Phi$) within an ODT of strontium monofluoride (SrF) by determining optimal polarizations of both the cooling and trapping light.  We have produced gases of up to $N_{ODT}\!\sim\!160$ trapped molecules with $T\!\sim\!14$\,$\mu$K, $n\!\sim\!2\times 10^{9}$\,cm$^{-3}$, and $\Phi\!\sim\!2\times 10^{-7}$.
	
	Our apparatus is illustrated in Fig.~\ref{fig:ODTSchematic}(a).  SrF molecules from a cryogenic buffer gas beam source~\cite{bsd2011} are slowed~\cite{bsd2012}, then cooled and trapped in an rfMOT~\cite{nmd2016,smd2016} ($N_{MOT}\!\sim \!3500$ molecules, $T_{MOT}\!\approx\!1$\,mK,  Gaussian width $\sigma_{MOT}\!\approx\!1$\,mm).  The rfMOT requires laser frequencies addressing all four $|X^{2}\Sigma^{+},v\!=\!0,N\!=\!1,J,F\rangle$ hyperfine levels (Fig.~\ref{fig:ODTSchematic}(b)), coupling them to the $|A^{2}\Pi_{1/2},v'\!=\!0,N'\!=\!0,J'\!=\!1/2,F'\rangle$ state, along with repumpers for the X($v\!=\!1,2,3$) vibrational states.  To use laser power efficiently, a single beam containing all needed frequencies is cycled through all three orthogonal axes of the rfMOT, then retro-reflected, to provide trapping and cooling along all axes.  To compensate for power loss due to optical elements along the path, an adjustable Keplerian telescope is used to control the convergence of the MOT beams. 
	
	Efficiently loading an ODT typically requires temperatures of $\lesssim 1/10$ the trap depth~\cite{ogt2001}, which is at most 570\,$\mu$K in our apparatus.  Ideally, the cooling method should be effective both inside and outside the trap volume.  One technique demonstrated to reach the required temperature in similar molecules (CaF and YO) requires coupling two hyperfine levels in the $|X^{2}\Sigma,N\!=\!1\rangle$ manifold to $|A^{2}\Pi_{1/2},J\!=\!1/2\rangle$, with an overall blue detuning $\Delta$ and a relative Raman detuning $\delta$ (Fig.~\ref{fig:ODTSchematic}(b)), to create a $\Lambda$ system~\cite{cad2018,dwy2020}.  This `$\Lambda$-cooling' approach combines gray molasses~\cite{twt2017} with the velocity-selective coherent population trapping (VSCPT) characteristic of $\Lambda$ systems~\cite{gfs2013,aac1988}, and is commonly used for loading alkali atoms into optical dipole traps~\cite{bvr2014}.  In CaF, coupling the $|F\!=\!2,J\!=\!3/2\rangle$ and $|F\!=\!1,J\!=\!1/2\rangle$ levels was found to cool molecules to $T\!\sim\!10$\,$\mu$K in free space~\cite{cad2018}.
	
	However, although SrF has a very similar level structure to CaF, we find that coupling the analogous pair of states in SrF does \textit{not} result in effective cooling.  Instead, we observe cooling to $T\!\sim\!10$\,$\mu$K in free space when coupling the $|F\!=\!1,J\!=\!3/2\rangle$ and $|F\!=\!1,J\!=\!1/2\rangle$ states for $\Delta/(2\pi)>9$\,MHz $ =1.4\Gamma$ (where $\Gamma/(2\pi)=6.63$\,MHz is the natural linewidth of the $X\!\rightarrow\!A$ transition).  So, in all the work described here we use those states for $\Lambda$-cooling (Fig.~\ref{fig:ODTSchematic}(b)).  This dependence on which state is coupled to $|F\!=\!1,J\!=\!1/2\rangle$ is also evident in a numerical simulation based on solving the Optical Bloch Equations (OBEs)~\cite{dta2018}.  For SrF, the simulation shows a much stronger damping force when using $|F\!=\!1,J\!=\!3/2\rangle$ than when using $|F\!=\!2,J\!=\!3/2\rangle$.  
	
	In CaF, $\Lambda$-cooling was also shown to be effective within the ODT~\cite{cad2018}.  This is remarkable, since---unlike alkali atoms in the $|^{2}\text{S}_{1/2}\rangle$ ground state---molecules in a $|^{2}\Sigma,N=1\rangle$ state have substantial vector and tensor polarizabilities even for far-detuned traps~\cite{kde2010}, which lead to differential shifts between the substates of each hyperfine manifold that can destabilize the zero-velocity dark states needed for effective VSCPT cooling~\cite{btg1995}.  For example, Zeeman shifts on the order of $\!\sim\!100$\,kHz ($\!\sim\!5\mu$K) were observed to limit sub-Doppler cooling of diatomic molecules in free space~\cite{cdt2019,twt2017,msd2018}.  The comparably large ODT-induced differential AC Stark shifts were thus proposed as an explanation for the observed saturation of trapping efficiency of CaF at a trap depth of 130\,$\mu$K, since higher trap depths lead to larger differential shifts~\cite{cad2018}.  
	
	We use $\Lambda$-cooling here for loading of SrF into an ODT. Unless otherwise indicated, here $\Delta/(2\pi) = +22$\,MHz; $\delta/(2\pi) = +1.2$\,MHz; the total intensity from all 6 passes of the $\Lambda$ laser beam, $I$, is 278\,mW/cm$^{2}$; and the ratio of intensities coupling the two hyperfine levels, $R_{1\uparrow,1\downarrow}$, where $1\!\uparrow\!$ ($1\!\downarrow\!$) denotes $|F\!=\!1,J\!=\!3/2\rangle$ ($|F\!=\!1,J\!=\!1/2\rangle$), is $R_{1\uparrow,1\downarrow}=2/3$.
	
	The ODT, formed by focusing a $\!\sim\!50$\,W single-mode 1064\,nm laser beam to a $1/e^{2}$-radius of $\!\sim$40\,$\mu$m, is turned on at the same time as the $\Lambda$-cooling.  This laser is combined with one $\Lambda$-cooling beam pass using a dichroic mirror, then passes through a $\lambda/4$ plate before entering the chamber.  We have measured the Jones matrix of those elements, which, along with the orientations of additional $\lambda/2$ and $\lambda/4$ waveplates (lower-left, Fig.~\ref{fig:ODTSchematic}(a)) in the ODT beam path prior to the dichroic, determines the polarization of the trap light. 
	
	The trap depth for a given ODT intensity is determined by calculating the AC Stark shift based on measured~\cite{dcz1974,ber1996} and/or calculated~\cite{NLEDM2019,dor2019} dipole matrix elements between the $|X^{2}\Sigma\rangle$ ground state and all excited states~\cite{supp}.  Then, by measuring the beam profile along the ODT axis ($z$), we determine the full axial trap depth profile $T_{D}(z)$ (Fig.~\ref{fig:ODTSchematic}(c)) and maximum trap depth $T_{D,0}=570\mu$K.  The profile deviates from the ideal quadratic behavior due to astigmatism of the ODT beam.
	
	After 150\,ms of simultaneous application of ODT and $\Lambda$ beams, the $\Lambda$-cooling light is shuttered for a time $t_{sh}$ ($50$\,ms unless otherwise indicated), to allow untrapped molecules to fall from the imaging region.  Then, $\Lambda$-cooling light is turned back on for 150\,ms, during which the camera is exposed.  Trapped molecules remain cold even as they scatter photons~\cite{cad2018}, allowing for their fluorescence to be imaged \textit{in situ} (Fig.~\ref{fig:ODTSchematic}(d)).  This indicates that $\Lambda$-cooling is effective inside the trap.  The peak trap-induced scalar AC stark shift for the $X$ state ($A$ state) is -11.9\,MHz (+0.4\,MHz) under our conditions.  These combine to redshift the one photon detuning in the trap center down to $\Delta_{trap}/(2\pi)=+10$\,MHz, which is still blue enough to cool effectively.  
	
	\begin{figure*}
		\includegraphics{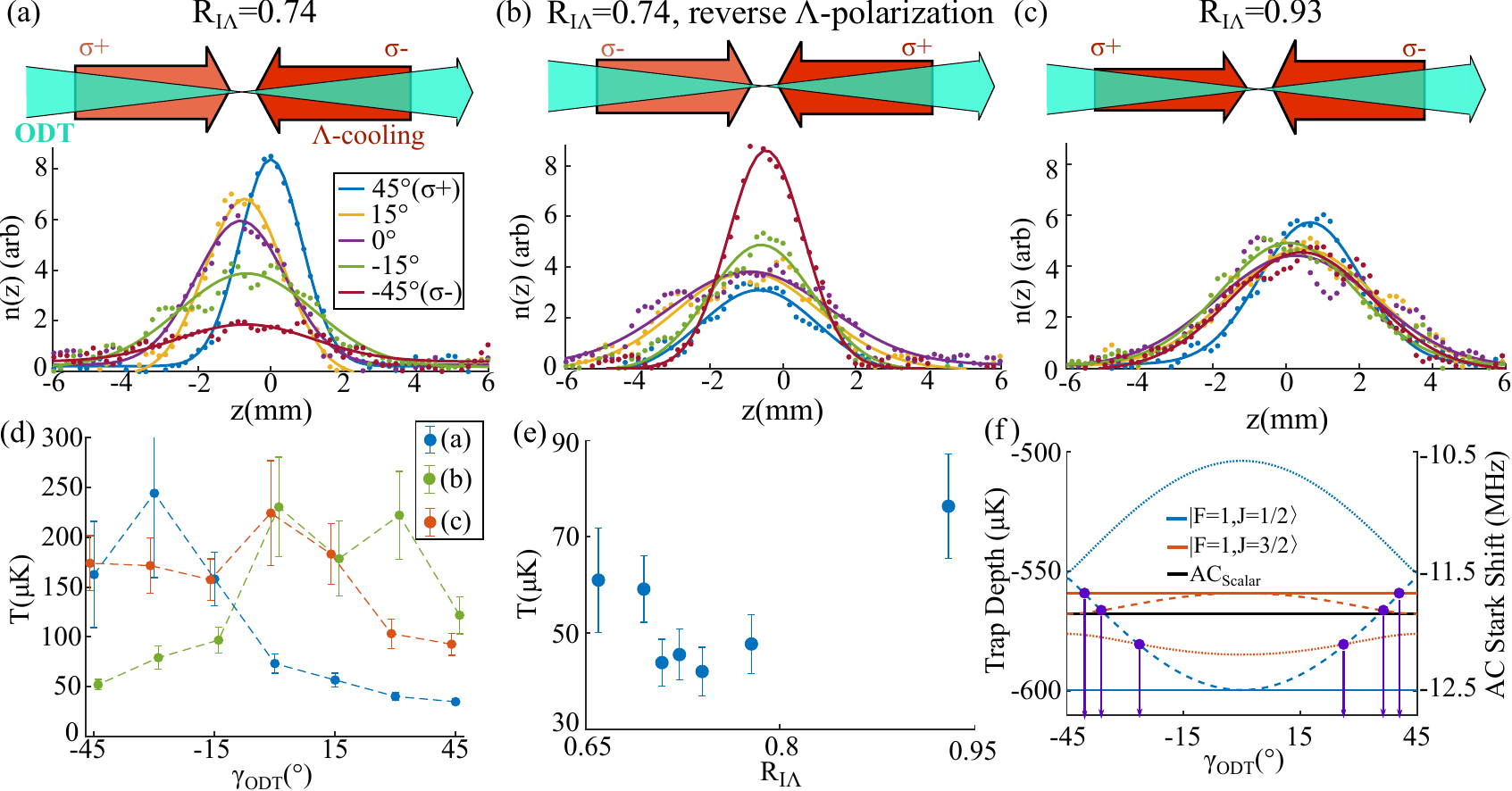}%
		\caption{(a)-(c) Trapped molecule profiles for different $R_{I\Lambda}$ (indicated by shading of $\Lambda$-cooling beam arrows) and $\gamma_{ODT}$ (in plot legends).  The trap is optimized when the ODT polarization matches that of the weaker of the two $\Lambda$ beams on its axis.  (a) With $R_{I\Lambda}=0.74$ and $\sigma^{+}$ polarization of the weaker $\Lambda$ beam, the trap is optimized for $\gamma_{ODT}=+45^{\circ}$.  (b) When the $\Lambda$-beam polarizations are reversed, $\gamma_{ODT}=-45^{\circ}$ is optimal. (c) If the imbalance is decreased so that $R_{I\Lambda}=0.93$ (by making the beam convergent), the dependence on $\gamma_{ODT}$ is reduced.  (d) $T$, as determined from fits to the profiles in (a-c), vs. $\gamma_{ODT}$. (e) $T$ vs $R_{I\Lambda}$ for $\gamma_{ODT}=+45^{\circ}$, with $\Lambda$ polarization as in (a).  (f) Peak trap depth (and associated AC Stark Shift) for each eigenstate of the ODT AC Stark Hamiltonian for the two $|X^{2}\Sigma,N\!=\!1,F=1\rangle$ states, including the tensor and vector polarizability, vs. $\gamma_{ODT}$~\cite{supp}.  Purple circles indicate values of $\gamma_{ODT}$ where a pair of states, one each from $F=1\!\uparrow$ and $F=1\!\downarrow$, are degenerate.\label{fig:polAndConverge}}
	\end{figure*}
	
	Our imaging resolution is insufficient to observe the molecular cloud width ($\sigma_{rad}$) along the short axes of the ODT, $\hat{x}$ and $\hat{y}$ (Fig.~\ref{fig:ODTSchematic}).  Due to the imbalance in trap frequencies $\omega_{x,y}\gg\omega_{z}$, the axial extent of the cloud, $\sigma_{ax}$, is large enough to be resolved by our camera, whose horizontal axis is at 45$^{\circ}$ with respect to the ODT axis ($\hat{z}$).  The cloud density profile along the ODT axis, $n(z)$, is determined by integrating the image along $y$ (Fig.~\ref{fig:polAndConverge}(a-c)).  With knowledge of both $T_{D}(z)$ and $\sigma_{ax}$, the temperature $T$ can be determined from $n(z)$.  We also measured $T$ through time-of-flight (TOF) expansion and found results consistent with, albeit less precise than, the temperatures measured from the \textit{in-situ} axial distribution.
	
	We expected that cooling inside the trap would be sensitive to the polarization of the trap laser, due to the differential AC Stark shifts~\cite{kde2010,nyj2012,gbc2017,lhw2021} in the $|X^{2}\Sigma\rangle$ state.  The largest of these shifts within the two coupled hyperfine manifolds are $\!\sim\!10$\% of the scalar (i.e., average) AC stark shift (Fig.~\ref{fig:polAndConverge}(f)). We anticipated that the optimum polarizations would be those for which two states, one within each of the $|1,3/2\rangle$ and $|1,1/2\rangle$ manifolds, experience the same AC stark shift (Fig.~\ref{fig:polAndConverge}(f)) and thus form a coherent dark state both inside and outside the trap. Since there is no other applied field to define a quantization axis, we expected the cooling to depend only on the aspect ratio of the ODT light polarization ellipse, and not on its orientation or rotation direction. 
	
	Instead, we found that the cooling can strongly depend on the direction of rotation of the polarization of the ODT beam.  To quantify this dependence, we define the ellipticity, $\gamma_{ODT}$, in terms of the dimensionless Stokes parameters ($S_{1,2,3}$) of the trap light~\cite{hecht}.  Specifically, $\gamma_{ODT}=\frac{1}{2}\tan^{-1}\frac{S_{3}}{\sqrt{S_{1}^{2}+S_{2}^{2}}}$.  The sign of $\gamma_{ODT}$ indicates the direction of rotation of the electric field vector (when viewing along the direction of light propagation), with $\gamma_{ODT}>0$ indicating clockwise, while $\tan(\gamma_{ODT})$ is the ratio of minor to major axes of the polarization ellipse.  
	
	We found that the symmetry between positive and negative $\gamma_{ODT}$ is broken by an intensity imbalance between the counter-propagating $\Lambda$-cooling beams, which have opposite circular polarizations.  Such an intensity imbalance can arise easily in our apparatus, where the $\Lambda$-cooling beam co-(counter)-propagating to the ODT beam is the last (first) pass of the long path of the retro-reflected $\Lambda$-cooling beam through the trapping chamber (Fig.~\ref{fig:ODTSchematic}(a)). For example, if the $\Lambda$-cooling beam is collimated then, due to losses through optical elements on the path, the final pass has only 74\% of the first pass intensity.  The ratio of intensities of final to first pass, $R_{I\Lambda}$, can be increased (decreased) by making the beam mildly convergent (divergent).  
	
	Data showing the effect of this broken symmetry on the in-trap molecule temperature is shown in Fig.~\ref{fig:polAndConverge}.  Fig.~\ref{fig:polAndConverge}(a) shows the results for a collimated beam ($R_{I\Lambda}=0.74$) when the $\Lambda$ beam co-propagating with the ODT is $\sigma^{+}$ polarized while the counter-propagating beam is $\sigma^{-}$ polarized.  We find that the molecular cloud width, and thus $T$, is minimized when the ODT polarization matches the circularity of the weaker, co-propagating beam ($\gamma_{ODT}=+45^{\circ}$).  This remains the case when the $\Lambda$ polarizations are reversed (Fig.~\ref{fig:polAndConverge}(b)), in which case $T$ is optimized for $\gamma_{ODT}=-45^{\circ}$.  If the $\Lambda$ beam is adjusted such that the intensity is better balanced (Fig.~\ref{fig:polAndConverge}(c)), the dependence of $T$ on $\gamma_{ODT}$ is much less pronounced (Fig.~\ref{fig:polAndConverge}(d)).  Ultimately, we find that $T$ is globally minimized when the $\Lambda$ beam intensities are deliberately imbalanced---in particular, when $R_{I\Lambda}=0.74$ and $\gamma_{ODT} = +45^{\circ}$ (Fig.~\ref{fig:polAndConverge}(e)) for the $\Lambda$ polarizations in Fig.~\ref{fig:polAndConverge}(a).  This configuration is used throughout the rest of this paper.  (As expected, we observe no dependence of $T$ on the trap polarization ellipse orientation angle $\psi=\frac{1}{2}\tan^{-1}\frac{S_{2}}{S_{1}}$.)
	
	In an attempt to understand this unanticipated dependence of $T$ on $\gamma_{ODT}$, we developed an OBE solver~\cite{dta2018} with the capability to include intensity imbalanced, retro-reflected beams.  We explicitly add the AC Stark Hamiltonian from the ODT light (including vector and tensor shifts), while AC Stark shifts from the imbalanced $\Lambda$ beams (which can be of comparable magnitude to those from the ODT laser under our conditions~\cite{supp}) are included implicitly in the OBEs.  This solver was benchmarked using results from comparable solvers~\cite{dta2018,dta2016} and also against experimental observations, such as rfMOT trap temperature~\cite{nmd2016} and capture velocity, $\Lambda$-cooling~\cite{cad2018}, and single frequency cooling~\cite{cdt2019}.  However, we were not able to reproduce the effects shown in Fig.~\ref{fig:polAndConverge}.  So, the mechanism behind the observed interplay between ODT polarization and $\Lambda$-beam intensity imbalance remains an open question.  
	\begin{figure}
		\includegraphics{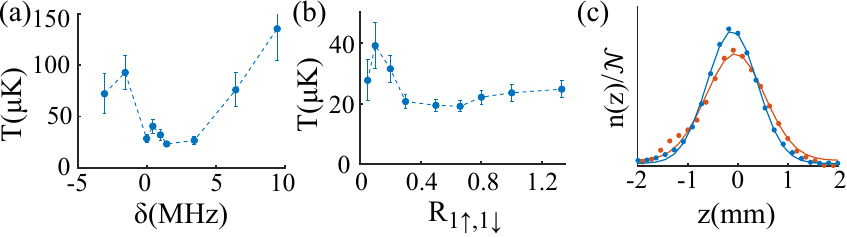}%
		\caption{Dependence of in-trap temperature on $\Lambda$ cooling parameters.  (a)  $T$ vs 2-photon detuning $\delta$, for fixed hyperfine ratio $R_{1\uparrow,1\downarrow}=2/3$.  (b) $T$ vs $R_{1\uparrow,1\downarrow}$ for fixed $\delta/(2\pi)=1.2$\,MHz.  (c) Examples of typical profile $n(z)/\mathcal{N}$, where $\mathcal{N}=\int{n(z)dz}$, with (blue, $T\!=\!14\mu$K) and without (red, $T\!=\!20\mu$K) fine tuning the spatial alignment of $\Lambda$-cooling beams. \label{fig:dependenceOnRamanParams}}
	\end{figure}
	
	We next worked to optimize $\delta$ and $R_{1\uparrow,1\downarrow}$.  In Fig.~\ref{fig:dependenceOnRamanParams}(a), we observe that $T$ is optimized near two-photon resonance ($\delta\!=\!0$) with a broad minimum extending to $\delta\!>\!0$.  Similar behavior has been observed in other $\Lambda$ cooling experiments~\cite{cad2018,gfs2013,rbm2018}.  The breadth of this feature is comparable to the in-trap two photon Rabi frequency between the coupled hyperfine manifolds ($\Omega_{\Lambda}/(2\pi)\!\approx\!8$\,MHz~\cite{supp}), as expected.  We also observe that $T$ is optimized for a hyperfine intensity ratio $R_{1\uparrow,1\downarrow}\!\approx\!2/3$ (Fig~\ref{fig:dependenceOnRamanParams}(b)), and becomes inefficient for $R_{1\uparrow,1\downarrow}\le0.25$.  We note in particular that `single frequency cooling' (where $R_{1\uparrow,1\downarrow}=0$)---which was shown to lead to $T\!<\!10\mu$K in free space for CaF~\cite{cdt2019}---is ineffective at cooling SrF in the ODT, though it performed as well as optimized $\Lambda$ cooling in free space.  
	
	For the optimal values of $\delta_{R}$ and $R_{1\uparrow,1\downarrow}$, we find that $T\!\sim\!20\mu$K is regularly achievable.  However, we have observed that $T$ is sensitively dependent on the spatial alignment of the $\Lambda$-cooling beams.  By iteratively adjusting the alignment while optimizing for $T$, we were able to achieve a minimum of $T=14(1)\mu$K (Fig.~\ref{fig:dependenceOnRamanParams}c).  Because this optimal condition was difficult to maintain, the SrF cloud had the more typical temperature in the data shown throughout this paper.
	
	\begin{figure}
		\includegraphics{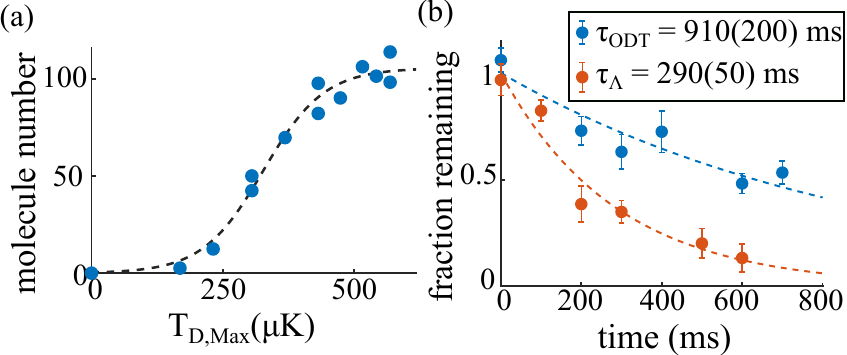}%
		\caption{(a) Number of molecules collected in ODT vs trap depth for optimum trap parameters ($\delta/(2\pi)=1.2$\,MHz, $\Delta/(2\pi)=22$\,MHz, $R_{1\uparrow,1\downarrow}=2/3$).  For reference, the rfMOT typically contains $\!\sim$3000 molecules.  The black dashed curve is a smooth curve used to guide the eye. (b) Fraction of molecules remaining vs time when $\Lambda$-cooling light is applied (red) and when it is shuttered (blue).  Dashed lines are exponential fits, with fit lifetime indicated in the legend. \label{fig:numVsODT}}
	\end{figure}

	Next, we studied the dependence of number of trapped molecules, $N_{ODT}$, on trap depth (Fig.~\ref{fig:numVsODT}(a)).  This is important because we want to capture as many molecules as possible, so we must ensure that our trap is deep enough to saturate $N_{ODT}$.  Furthermore, it was observed that the trap loading efficiency of CaF molecules peaked for $T_{D}\!\sim\!130$\,$\mu$K~\cite{cad2018}, and thus we wanted to check if SrF loading had similar behavior.  For this measurement, we recapture and image the optically trapped molecules in an rfMOT.  This is done by turning on the rfMOT coils and switching from the $\Lambda$-cooling laser configuration to the rfMOT configuration.  After the ODT is loaded, but prior to the rfMOT recapture, we turn off all cooling light for $t_{sh}=140$\,ms to ensure that untrapped molecules fall out of the MOT capture volume and are not detected.
	
	We find that $N_{ODT}$ rises monotonically with trap depth, but appears to saturate for $T_{D}\gtrsim500$\,$\mu$K.  We suspect that this striking difference in behavior, relative to CaF, relates to which states are chosen for $\Lambda$-cooling.  Hyperfine induced mixing between $|1,1/2\rangle$ and $|1,3/2\rangle$ modifies the transition strengths from the $|A\Pi_{1/2}\rangle$ hyperfine states to these levels.  In SrF, the $|1,3/2\rangle$ state couples 56$\times$ more strongly to $|A^{2}\Pi_{1/2},J\!=\!1/2,F\!=\!0\rangle$ than to $|A^{2}\Pi_{1/2},J\!=\!1/2,F\!=\!1\rangle$~\cite{supp}, so we expect that the $\Lambda$ system primarily couples through the former.  This avoids any complications that may arise due to the large vector light shift in the $|A^{2}\Pi_{1/2},J\!=\!1/2,F\!=\!1\rangle$ state~\cite{supp}.  If the $|2,3/2\rangle$ and $|1,1/2\rangle$ states are used, as in CaF~\cite{cad2018}, the coupling must be through $|A^{2}\Pi_{1/2},J\!=\!1/2,F\!=\!1\rangle$.  The larger number of sublevels, all of which experience differential shifts, in the latter scheme may also limit the effectiveness of $\Lambda$-cooling in deeper ODTs. 
	
	Another critical trap quantity is the lifetime $\tau_{ODT}$.  In order to study SrF collisions, we will need $\tau_{ODT}^{-1}\lesssim\ \beta n$, where $\beta$ is a collisional rate coefficient.  We measure $\tau_{ODT}$ by shuttering the $\Lambda$-cooling light for a variable time before re-opening the shutter and imaging the remaining trapped molecules.  We find $\tau_{ODT}=910(200)$\,ms (Fig.~\ref{fig:numVsODT}(b)).  Since this is comparable to the lifetime we measure in a magnetic quadrupole trap with the same background pressure~\cite{msd2018}, it seems likely that the lifetime in both cases is limited by collisions with background gas.
	
	We also measure the lifetime of molecules in the ODT while $\Lambda$-cooling light is applied, $\tau_{\Lambda}$.  This quantity sets the time over which molecule loading is effective and over which \textit{in-situ} imaging can occur.  To measure $\tau_{\Lambda}$, we continuously apply the $\Lambda$ light for an additional $t_{c}=650$\,ms after the 150\,ms loading and 50\,ms release times, and image for 50\,ms intervals during $t_{c}$.  
	
	We measure $\tau_{\Lambda}=290(50)$\,ms, similar to what was observed in an ODT of CaF~\cite{cad2018}.  In that experiment, the increased loss observed in presence of $\Lambda$ light was attributed to spatial diffusion out of the trap induced by scattering events.  However, Monte Carlo simulations indicate that this should contribute negligibly to the SrF loss rate at our ratio of $T_{D}/T$.  Light-assisted collisions represent another potential loss mechanism.  $\Lambda$-light assisted loss rate coefficients of $\beta\gtrsim10^{-9}$\,cm$^{3}$/s have been observed in diatomic molecules held in optical tweezers~\cite{acd2019}, so this effect could plausibly limit lifetimes to the few 100\,ms level for our typical peak density of $10^{9}$\,cm$^{-3}$.
	
	For applications where high-fidelity detection is critical, such as studying molecules prepared in arrays of optical tweezers~\cite{cad2020,acd2019}, it is important to scatter large numbers of photons per molecule.  The detection efficiency increases with the average number of photons emitted per molecule during one imaging lifetime, $t_{\Lambda}R_{\Lambda}$, where $R_{\Lambda}$ is the scattering rate during $\Lambda$-cooling.  We measure $R_{\Lambda}=3.1\times 10^{5}$s\,$^{-1}$ by comparing the fluorescence collected with that from the MOT recapture, where the scattering rate is known~\cite{nmd2016}.  Thus, $t_{\Lambda}R_{\Lambda}\!=\!9(2)\times 10^{4}$, $\sim3$ times larger than demonstrated in an ODT of CaF~\cite{cad2018} despite the similar $t_{\Lambda}$.  We attribute the larger $R_{\Lambda}$ observed here to the smaller in-trap detuning ($\Delta_{trap}/\Gamma=1.5$ here compared to $\Delta_{trap}/\Gamma=3.6$ in~\cite{cad2018}).
	
	In conclusion, by optimizing the combination of trap light polarization and intensity imbalance of $\Lambda$-enhanced gray molasses lasers, we have loaded $\!\sim\!5$\% of SrF molecules from our rfMOT into a $T_{D}\!=\!570\mu$K deep ODT, at temperatures as low as $T\!=\!14(1)\mu$K.  The large value of the ratio $T_{D}/T$ implies strong compression of the molecular cloud, yielding density and phase space density higher than previously reported in bulk gases of directly cooled molecules, despite starting with 10 times fewer molecules.
	
	We find that several features of loading molecule ODTs using $\Lambda$-cooling remain poorly understood, such as the dependence on the ratio of $\Lambda$ light intensities, the observed interplay between the trap polarization and cooling light intensity imbalance, and, more generally, the effect of vector and tensor light shifts.  Once these features are better understood, it is possible that higher trap compression could be achieved.  
	
	We are currently working on ways to increase both the number of molecules in our rfMOT and the loading efficiency into the ODT.  For the high compression achieved in our trap, a factor of 5 increase in $N_{ODT}$ would result in a universal-rate collision loss rate~\cite{iju2010}, $\tau_{0}\!=\!(\beta_{0}n)^{-1}\!\sim\!\tau_{ODT}$. This would be sufficient to allow collisions of ultracold SrF molecules to be studied for the first time.

	\nocite{swb2009,esc1983,nsc1988,wat2008,wkt2008,asa1968}
	
	\begin{acknowledgments}
		We gratefully acknowledge support from AFOSR, ONR, ARO (DURIP), and the Yale Quantum Institute.  We thank Loic Anderegg for helpful discussions.
	\end{acknowledgments}

\end{document}


	
	\title{Supplemental Material: Polarization Enhanced Deep Optical Dipole Trapping of $\Lambda$-cooled Polar Molecules}
	
	
	\author{Thomas K. Langin}
	\affiliation{Department of Physics, Yale University, New Haven, Connecticut, CT, 06520, USA}
	\affiliation{Yale Quantum Institute, Yale University, New Haven, Connecticut 06520, USA}
	\author{Varun Jorapur}
	\affiliation{Department of Physics, Yale University, New Haven, Connecticut, CT, 06520, USA}
	\affiliation{Yale Quantum Institute, Yale University, New Haven, Connecticut 06520, USA}
	\author{Yuqi Zhu}
	\affiliation{Department of Physics, Yale University, New Haven, Connecticut, CT, 06520, USA}
	\affiliation{Yale Quantum Institute, Yale University, New Haven, Connecticut 06520, USA}
	\author{Qian Wang}
	\affiliation{Department of Physics, Yale University, New Haven, Connecticut, CT, 06520, USA}
	\affiliation{Yale Quantum Institute, Yale University, New Haven, Connecticut 06520, USA}
	\author{David DeMille}
	\affiliation{Department of Physics, University of Chicago, Chicago, Illinois 60637, USA}
	
	\date{\today}
	
	
	\maketitle
	
	\section{Calculating AC Stark Shifts}
	
	We determine the AC Stark Hamiltonian resulting from monochromatic light of intensity $I$ and frequency $\omega_{L}$ at a given polarization $\hat{p}$ through second order perturbation theory.  We find that the matrix elements of the effective AC Stark Hamiltonian for two sublevels $|i_{1}\rangle$, $|i_{2}\rangle$ within the same rovibronic state $|i\rangle$ can be written as:
	\begin{equation}
		\langle i_{1}|H_{AC}|i_{2}\rangle=-\frac{I}{2c\hbar\epsilon_{0}}\sum_{f}\left(\frac{d_{fi}^{2}}{\omega_{fi}-\omega_{L}}+\frac{d_{fi}^{2}}{\omega_{fi}+\omega_{L}}\right)\langle i_{1}|\hat{r}\cdot\hat{p}|f\rangle\langle f|\hat{r}\cdot\hat{p}|i_{2}\rangle
		\label{eq:ACStark}
	\end{equation}
	
	\noindent where we consider specifically the AC Stark Hamiltonians for two values of $|i\rangle$; $|i\rangle=|X^{2}\Sigma,v=0,N=1\rangle$ and $|i\rangle=|A^{2}\Pi_{1/2},v=0,J=1/2\rangle$, $f$ refers to all other states, $\omega_{fi}=\omega_{f}-\omega_{i}$ where $\hbar\omega_{f}$ ($\hbar\omega_{i}$) are the energies of state $|f\rangle$ ($|i\rangle$), $d_{fi}=\langle f|er|i\rangle$ is the transition dipole matrix element, and we have included the co-rotating term.  In Table~\ref{tb:transMatrix} we list the molecule-frame transition dipole matrix elements we use for calculating the AC Stark shift.
	
	\begin{table}
		\begin{tabular}{ l | c | c | r }
			$|i\rangle$ & $|f\rangle$ & $d_{fi}$ (Debye) & $E_{f}-E_{i}$ (cm$^{-1}$)\\ \hline
			$|X^{2}\Sigma\rangle$ & $|A^{2}\Pi_{1/2}\rangle$ & \textbf{6.22}~\cite{dcz1974} & \textbf{15076}~\cite{swb2009}\\ 
			$|X^{2}\Sigma\rangle$ & $|A^{2}\Pi_{3/2}\rangle$ & \textbf{6.24}~\cite{dcz1974} & \textbf{15357}~\cite{swb2009}\\ 
			$|X^{2}\Sigma\rangle$ & $|B^{2}\Sigma_{1/2}\rangle$ & \textbf{4.93}~\cite{ber1996} & \textbf{17267}~\cite{esc1983} \\ 
			$|X^{2}\Sigma\rangle$ & $|C^{2}\Pi_{1/2}\rangle$\&$|C\Pi_{3/2}\rangle$ & 1.53~\cite{dor2019} & \textbf{27385}~\cite{swb2009}\\ 
			$|X^{2}\Sigma\rangle$ & $|D^{2}\Sigma_{1/2}\rangle$ & 0.82~\cite{dor2019} & \textbf{27774}~\cite{swb2009} \\ 
			$|X^{2}\Sigma\rangle$ & $|F^{2}\Sigma_{1/2}\rangle$ & 1.02~\cite{dor2019} & \textbf{32824}~\cite{nsc1988} \\ 
			$|X^{2}\Sigma\rangle$ & $|G^{2}\Pi_{1/2}\rangle$\&$|G\Pi_{3/2}\rangle$ & 1.27~\cite{dor2019} & \textbf{34809}~\cite{nsc1988}\\ 
			$|A^{2}\Pi_{1/2}\rangle$ & $|A^{\prime2}\Delta_{3/2}\rangle$ & 6.89~\cite{NLEDM2019} & 4036~\cite{NLEDM2019} \\ 
			$|A^{2}\Pi_{1/2}\rangle$ & $|B^{2}\Sigma_{1/2}\rangle$ & 0.53~\cite{NLEDM2019} & \textbf{2195}~\cite{esc1983} \\ 
			$|A^{2}\Pi_{1/2}\rangle$ & $|C^{2}\Pi_{1/2}\rangle$\&$|C\Pi_{3/2}\rangle$ & 2.80~\cite{dor2019} & \textbf{12313}~\cite{swb2009}\\ 
			$|A^{2}\Pi_{1/2}\rangle$ & $|D^{2}\Sigma_{1/2}\rangle$ & 5.34~\cite{dor2019} & \textbf{12702}~\cite{swb2009}\\ 
		\end{tabular}
		\caption{\label{tb:transMatrix}Transition dipole matrix elements used in AC Stark shift calculations.  Bold values indicate experimental results; the others are theoretical calculations.}
	\end{table}
	
	\subsection{AC Stark shift from ODT for $|i\rangle=|X^{2}\Sigma,N=1\rangle$}
	
	First we consider the case where $|i\rangle=|X^{2}\Sigma,N=1\rangle$.\\
	
	To calculate the $\langle f|\hat{r}\cdot\hat{p}|i\rangle$ terms when $f$ is a $|\Pi\rangle$ state (e.g. Hund's case (a)) we first express $|X^{2}\Sigma,N=1\rangle$ (Hund's case (b): $|F',N',J',m_{F}'\rangle$) in the Hund's case (a) basis~\cite{wat2008}
	
	\begin{equation}
		|\Lambda;N,S,J\rangle =\sum_{\Omega=-1/2}^{1/2}\sum_{\Sigma=-1/2}^{1/2}(-1)^{J+\Omega}\sqrt{2N+1}\begin{pmatrix} S&N&J\\ \Sigma&\Lambda&-\Omega\end{pmatrix}|\Lambda,S,\Sigma,\Omega,J\rangle
	\end{equation}
	
	\noindent We then follow the method in the appendix of~\cite{wkt2008} to evaluate $|\langle f|\hat{r}\cdot\hat{p}|i\rangle|^{2}$ for a particular set of quantum numbers \{$F'$,$J'$,$m_{F}'$\} for state $|i\rangle$.\\  
	
	When $|f\rangle$ is a $|\Sigma\rangle$ (case (b) ($|f;F,N,J,m_{F}\rangle$)) state, we determine:
	
	\begin{equation}
		|\langle f|\hat{r}\cdot\hat{p}|i\rangle|=m_{1}m_{2}m_{3}m_{4}
	\end{equation}
	
	\noindent where
	
	\begin{equation}
		\begin{aligned}
			m_{1} &= (-1)^{F-m_{F}}\begin{pmatrix}F&1&F'\\-m_{F}&-p&m_{F}'\end{pmatrix}\\
			m_{2} &= (-1)^{F'+J+1+I}\sqrt{(2F+1)(2F'+1)}\begin{Bmatrix}J'&F'&I\\F&J&1\end{Bmatrix}\\
			m_{3} &= (-1)^{J'+N+1+S}\sqrt{(2J+1)(2J'+1)}\begin{Bmatrix}N'&J'&S\\J&N&1\end{Bmatrix}\\
			m_{4} &= (-1)^{N}\sqrt{(2N+1)(2N'+1)}\begin{pmatrix}N&1&N'\\0&0&0\end{pmatrix}
		\end{aligned}
	\end{equation}
	\noindent and $p$ refers to a polarization component of the ODT, $p=\{0,\pm1\}=\{\pi,\sigma^{\pm}\}$.\\
	
	In Table~\ref{tb:shiftsCircPol}, we show the solution of Eq.~\ref{eq:ACStark} for $\hat{p}=\sigma^{+}$ for the allowed values of $|X^{2}\Sigma;N=1,F,m_{F}\rangle$ for $I=2.08$\,MW/cm$^{2}$ (the peak intensity of the ODT).  We note here that, throughout, we write energies $E$ in units either of angular frequency ($\omega=E/\hbar$), temperature ($T=E/k_{B}$), or wavenumber ($k=E/(2\pi\hbar c)$).\\
	
	\begin{table}
		\begin{tabular}{ l | c | c |c | r }
			State ($|F,m_{F}\rangle$) & $E_{AC}/(2\pi)$ (MHz) ($T_{D}$ ($\mu$K)) & $\Delta E_{AC}/(2\pi)$ (kHz) ($\Delta T_{D}$ ($\mu$K)) & $\frac{\alpha_{V}}{\alpha_{S}}$ & $\frac{\alpha_{T}}{\alpha_{S}}$\\ \hline
			$|1\downarrow,-1\rangle$ & -11.52 (553) & 309 (15) & ~ & ~\\
			$|1\downarrow,0\rangle$ & -12.49 (599) & -665 (-32) & -0.0020 & 0.0563 \\
			$|1\downarrow,1\rangle$ & -11.47 (550) & 356 (17) & ~ & ~\\
			$|0,0\rangle$ & -11.83 (568) & 0 (0) & ~ & ~\\
			$|1\uparrow,-1\rangle$ & -12.00 (576)& -177 (8) & ~ & ~\\
			$|1\uparrow,0\rangle$ & -11.64 (558) & 178 (8) & -0.0074 & -0.0151\\
			$|1\uparrow,1\rangle$ & -11.83 (568) & -2 ($\approx\!0$) & ~ & ~\\
			$|2,-2\rangle$ & -11.38 (546) & 442 (21) & ~ & ~\\
			$|2,-1\rangle$ & -12.09 (580) & -266 (-13) & ~ & ~ \\
			$|2,0\rangle$ & -12.31 (591) & -487 (-23) & -0.0038 & 0.0824\\
			$|2,1\rangle$ & -12.05 (578) & -221 (-11) & ~ & ~\\
			$|2,2\rangle$ & -11.29 (542) & 531 (26) & ~ & ~\\
		\end{tabular}
		\caption{\label{tb:shiftsCircPol} First column indicates the $|F,m_{F}\rangle$ states in the $|X^{2}\Sigma\rangle$ level.  As in the main text, $\downarrow$ refers to $|F=1,J=1/2\rangle$ and $\uparrow$ to $|F=1,J=3/2\rangle$.  Second column is the AC Stark shift (and associated trap depth) for each level for $\sigma^{+}$ polarized ODT light (with wavelength $\lambda=1064$\,nm) with $I\!\sim\!2.0$\,MW/cm$^{2}$.  Third column is $E-E_{s}$ where $E_{s}$ is the scalar Stark shift.  Fourth and fifth columns are the extrapolated vector ($\alpha_{V}$) and tensor ($\alpha_{T}$) polarizabilities normalized by the scalar polarizability $\alpha_{S}$.  Since these values correspond to a given hyperfine manifold (as opposed to individual $|m\rangle$ states), I display the value for each manifold next to the corresponding $|m=0\rangle$ sub-level.}
	\end{table}

	The AC Stark Hamiltonian can be written in terms of a scalar ($\alpha_{s}$), vector ($\alpha_{v}$), and tensor ($\alpha_{T}$) polarizability~\cite{asa1968}.  For an $|F=1\rangle$ state this is, in the $|F,m_{F}\rangle$ basis
	
	\begin{equation}
		\begin{split}
			\frac{H_{Stark}}{E_{s}}&=\begin{pmatrix}1&0&0\\0&1&0\\0&0&1\end{pmatrix}-\frac{\alpha_{V}}{\alpha_{S}}\begin{pmatrix}-\sin(2\gamma_{ODT})&0&0\\0&0&0\\0&0&\sin(2\gamma_{ODT})\end{pmatrix}\\
			&-\frac{\alpha_{T}}{\alpha_{S}}\begin{pmatrix}\frac{1}{2}&0&\frac{3}{2}\cos(2\gamma_{ODT})\\0&-1&0\\\frac{3}{2}\cos(2\gamma_{ODT})&0&\frac{1}{2}\end{pmatrix}
		\end{split}
		\label{eq:HStarkPolarF1}
	\end{equation}
	
	\noindent where $E_{s}$ is the scalar Stark shift (equal to the shift for $|0,0\rangle$, and also to the average shift over all $|F,m\rangle$ states, in Table~\ref{tb:shiftsCircPol}).  Even though the two $|F=1\rangle$ states are the ones that we couple in our $\Lambda$-system, the molecule has a non-trivial probability of occupying the $|F=2\rangle$ state during the cooling process, and so the Stark shifts in that level may also play some role in how the cooling performs.  For $|F=2\rangle$, we find:
	
	\begin{equation}
		\begin{split}
			\frac{H_{Stark}}{E_{s}}&=\begin{pmatrix}1&0&0&0&0\\0&1&0&0&0\\0&0&1&0&0\\0&0&0&1&0\\0&0&0&0&1\end{pmatrix}-\frac{\alpha_{V}}{\alpha_{S}}\begin{pmatrix}-\sin(2\gamma_{ODT})&0&0&0&0\\0&-\frac{\sin(2\gamma_{ODT})}{2}&0&0&0\\0&0&0&0&0\\0&0&0&\frac{\sin(2\gamma_{ODT})}{2}&0\\0&0&0&0&\sin(2\gamma_{ODT})\end{pmatrix}\\
			& -\frac{\alpha_{T}}{\alpha_{S}}\begin{pmatrix}\frac{1}{2}&0&\sqrt{\frac{3}{8}}\cos(2\gamma_{ODT})&0&0\\0&-\frac{1}{4}&0&\frac{3}{4}\cos(2\gamma_{ODT})&0\\\sqrt{\frac{3}{8}}\cos(2\gamma_{ODT})&0&-\frac{1}{2}&0&\sqrt{\frac{3}{8}}\cos(2\gamma_{ODT})\\0&\frac{3}{4}\cos(2\gamma_{ODT})&0&-\frac{1}{4}&0\\0&0&\sqrt{\frac{3}{8}}\cos(2\gamma_{ODT})&0&\frac{1}{2}\end{pmatrix}
		\end{split}
		\label{eq:HStarkPolarF2}
	\end{equation}
	
	For $p=\pm1$ ($\gamma_{ODT}=\pm45$) these Hamiltonians are diagonal.  So, the ratios of $\alpha_{V}/\alpha_{S}$ and $\alpha_{T}/\alpha_{S}$ can be determined directly from column 2 of Table.~\ref{tb:shiftsCircPol}, and are displayed in the fourth and fifth columns of that table.  
	
\begin{figure}
		\includegraphics{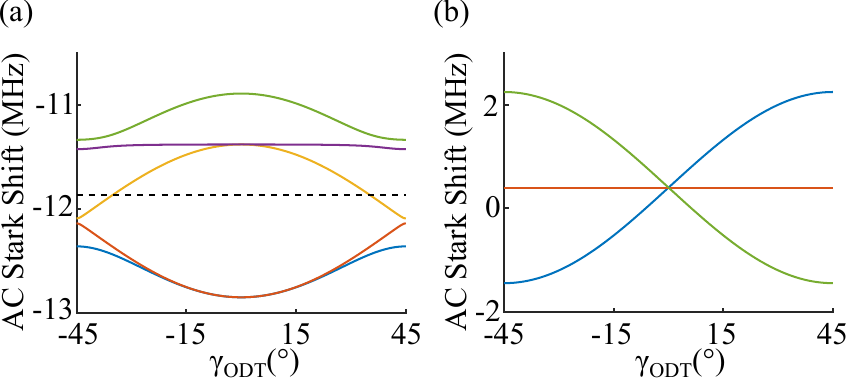}%
		\caption{(a) Solid lines are eigenenergies of the AC Stark Hamiltonian in the $|X^{2}\Sigma,N=1,F=2\rangle$ manifold from solving Eq.~\ref{eq:HStarkPolarF2} for $\lambda$=1064\,nm and $I=2.0$\,MW/cm$^{2}$ as a function of $\gamma_{ODT}$ for the determined values of $\frac{\alpha_{T}}{\alpha_{S}}$ and $\frac{\alpha_{V}}{\alpha_{S}}$ in the $F=2$ level.  Black dashed line is $E_{s}$.  (b) Eigenenergies of the AC Stark Hamiltonian for the $|A^{2}\Pi_{1/2},J=1/2\rangle$ state under the same conditions as (a).  Because the tensor polarizability $\alpha_{T}=0$ in this state, the Hamiltonian is diagonal in the $|F,m_{F}\rangle$ basis.  For (b), blue corresponds to $|F=1,m_{F}=+1\rangle$, green to $|F=1,m_{F}=-1\rangle$, and red to both $|F=0,m_{F}=0\rangle$ and $|F=1,m_{F}=0\rangle$. \label{fig:suppACStark}}
	\end{figure}

	
	Plugging those values into Eq.~\ref{eq:HStarkPolarF1} and Eq.~\ref{eq:HStarkPolarF2} and solving for the eigenvalues as a function of $\gamma_{ODT}$, we find the curves shown in Fig.~2(f) in the main text for the two $|F=1\rangle$ states, along with those in Fig.~\ref{fig:suppACStark}(a) for the $|F=2\rangle$ state.
	
	\subsection{AC Stark shift from ODT beams for $|i\rangle=|A^{2}\Pi_{1/2},J=1/2\rangle$}
	
	The ODT also induces AC Stark shifts in the $|A^{2}\Pi_{1/2},J=1/2\rangle$ state.  In Table~\ref{tb:transMatrix} we show the additional transition dipole matrix elements needed to calculate the AC Stark shifts for this level.  By calculating $\alpha_{S}$ and $\alpha_{V}$ ($\alpha_{T}=0$ for this state) in the same manner as we did above for $|X^{2}\Sigma,N=1\rangle$, we can determine the eigenenergies of the AC Stark matrix in this state vs $\gamma_{ODT}$ (Fig.~\ref{fig:suppACStark}(b)).
	
	\subsection{AC Stark shift from $\Lambda$-cooling beams for $|i\rangle=|X^{2}\Sigma,N=1\rangle$}
	
	To estimate whether the differential AC Stark shifts resulting from the $\Lambda$-cooling lasers could be large enough to compensate for those from the ODT, we solve for the AC Stark shifts resulting from a 1D pair of counter-propagating cross-circularly-polarized $\Lambda$ beams with one photon detuning $\Delta/(2\pi)$ equivalent to the `in-trap' value (e.g. with the ODT induced Stark shifts for the X (Fig.~2(f) of the main text) and A (Fig.~\ref{fig:suppACStark}(b)) states added in) of 10\,MHz, $\delta_{R}/(2\pi)=1.2$\,MHz, $R_{1\uparrow,1\downarrow}=2/3$, $R_{I\Lambda}=0.74$, and the beam intensity (summed over both hyperfine addressing frequency components) of the initial (stronger) pass is 46\,mW/cm$^{2}$, matching the experimental conditions described in the paper. \\ 
	
	We found that $\langle F=1\downarrow,m_{F}=-1|H_{AC}/(2\pi)|1\downarrow,-1\rangle-\langle 1\downarrow,+1|H_{AC}/(2\pi)|1\downarrow,+1\rangle=205$\,kHz and $\langle 1\uparrow,-1|H_{AC}/(2\pi)|1\uparrow,-1\rangle-\langle 1\uparrow,+1|H_{AC}/(2\pi)|1\uparrow,+1\rangle=140$\,kHz ($m_{F}$ remains a good quantum number in the presence of only circularly polarized light).  These differential shifts are comparable to the ones induced by the ODT (Fig.~2(f) of the main text) when $\gamma_{ODT}=\pm45^{\circ}$, and so could conceivably play a role in either mitigating or enhancing the effect of differential intra-hyperfine manifold shifts on the in-trap gray-molasses cooling.
	
	\section{Differential Transition Strengths and Two Photon Rabi Frequency for $\Lambda$-cooling}
	
\begin{figure}
		\includegraphics{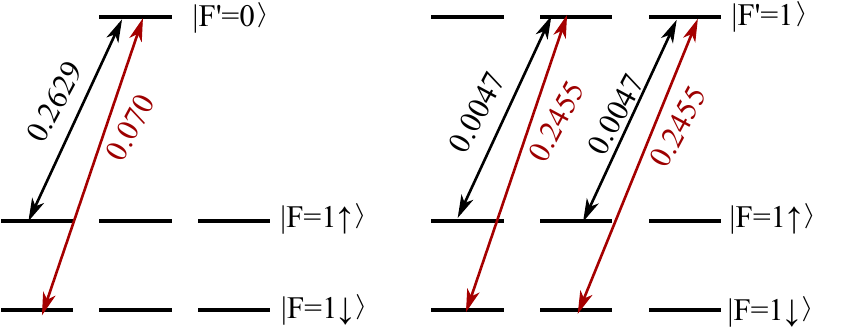}%
		\caption{$\Lambda$-coupling diagram for the $|A^{2}\Pi_{1/2},J=1/2,F'\rangle \leftrightarrow |X^{2}\Sigma,N=1,F,J\rangle$ transition for the case of $\sigma^{+}$ polarized light.  Values next to the arrows indicate $f_{F,F'}$.  Left: $F'=0$, Right: $F'=1$. \label{fig:branching}}
	\end{figure}

	In the main text we implied that the $\Lambda$-cooling light primarily couples the $|X\Sigma,F\!=\!1\!\uparrow\rangle$ and $|X\Sigma,F\!=\!1\!\downarrow\rangle$ states through the $|A\Pi,F'\!=\!0\rangle$ manifold and not the $|A\Pi,F'\!=\!1\rangle$ manifold.  Here, we justify that assumption by considering the simple scenario for when the two manifolds are coupled by $\sigma^{+}$ polarized light.  In Figure~\ref{fig:branching} we illustrate the branching ratios for decays from the $|A\Pi,F'\rangle$ manifold to the $|X\Sigma,F\rangle$ manifold, $f_{F,F'}$, which are proportional to the squares of transition dipole matrix elements, for states that are coupled by the $\sigma^{+}$ $\Lambda$-cooling light.
	
	

	The two photon Rabi frequency, for the case when the amplitude through a single excited state $F'$ dominates, is given by:
	
	\begin{equation}
		\Omega_{F'}=\frac{\Gamma^{2}}{4\Delta}\sqrt{f_{1\uparrow,F'}f_{1\downarrow,F'}}\sqrt{\frac{I_{1\uparrow,F'}I_{1\downarrow,F'}}{{I_{sat}^{2}}}}\label{eq:TwoPhotRabi}
	\end{equation}
	
	\noindent where $I_{F,F'}$ refers to the intensity of the light driving the transition between states $|X\Sigma,F\rangle$ and $|A\Pi,F'\rangle$, and $I_{sat}=\pi hc\Gamma/(3\lambda^{3})$ ($\lambda=663$\,nm for this transition) is the saturation intensity ($I_{sat}=2.9$\,mW/cm$^{2}$).  Inserting the $f_{F,F'}$ values shown in Fig.~\ref{fig:branching}, the in-trap overall detuning $\Delta/(2\pi)=10$\,MHz, and the intensity (278\,mW/cm$^{2}$) and hyperfine ratio $R_{1\uparrow,1\downarrow}$ (2/3), used in the experiment into Eq.~\ref{eq:TwoPhotRabi}, we find $\Omega_{0}/(2\pi)=8.5$\,MHz and $\Omega_{1}/(2\pi)=2.1$\,MHz for each of the two $F'=1$ coupled states, justifying the assumption that the hyperfine levels are primarily coupled through $F'=0$.  The disparity in two-photon Rabi frequencies is primarily due to the extraordinarily low branching ratio between $|A\Pi,F'=1\rangle$ and $|X\Sigma,F=1\uparrow\rangle$ of $f_{1\uparrow,1}=0.0047$.  We find that $\Omega_{0}/(2\pi)$ is similar to the width of the Raman resonance width observed in Fig.~3(a) of the main text.